\documentclass[12pt,a4paper]{article}
\usepackage{psfig}
\usepackage{array}

\title{Discrete Physics:\\
a new way to look at cryptography}
\author{
B. Chopard
\thanks{Computer Science Department, University of Geneva,  
1211, Geneve 4, E-mail: Bastien.Chopard@cui.unige.ch }\\
\centerline{and}\\
S. Marconi
\thanks{2 avenue de Frontenex, 1207 Geneve, E-mail: marconi@cui.unige.ch }
}
\date{}

\def\dt{\Delta t}
\def\v{\vec v}
\def\r{\vec r}
\def\Crypto{{\tt Crystal }}
\def\Finv{{\cal F}^{-1}}

\begin{document}

\maketitle

\begin{abstract}
This paper shows that Physics is very close to the
substitution-diffusion paradigm of symmetric ciphers. Based on this
analogy, we propose a new cryptographic algorithm. Statistical Physics
gives design principles to devise fast, scalable and secure encryption
systems. In particular, increasing space dimension and considering
larger data blocks improve both speed and security, allowing us to
reach high throughput (larger than 10Gb/s on dedicated HW).  The physical
approach enlarges the way to look at cryptography and is expected to
bring new tools and concepts to better understand and quantify
security aspects.
\end{abstract}

\section{Introduction}
\begin{figure}
\centerline{\psfig{figure=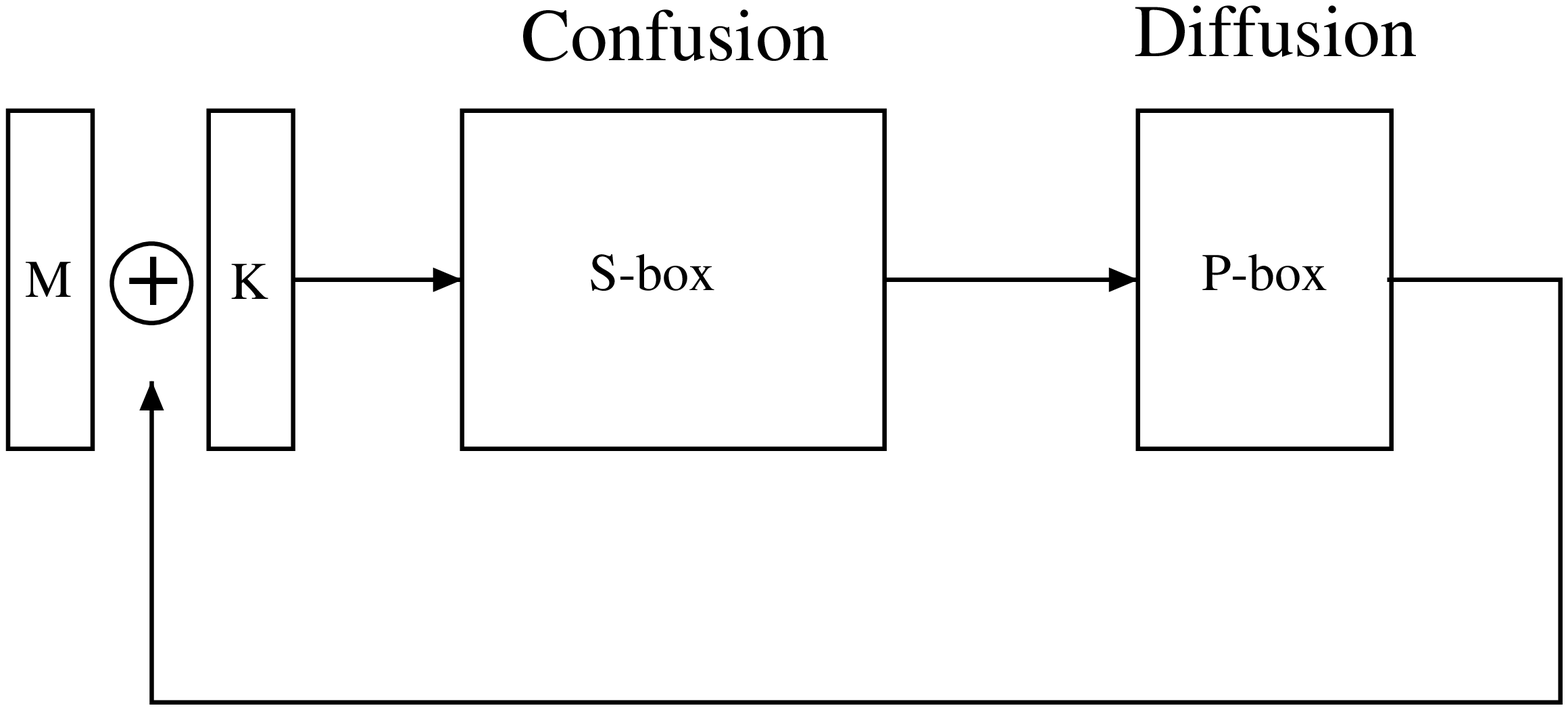,width=.6\textwidth}}
\caption{The standard structure of a symmetric block cipher.}
\label{fig:crytoStructure}
\end{figure}

Symmetric cryptography consists in transforming a plain text message
$M$ into a cipher text $M'$ through a series of operations that can
be inverted by the recipient of the encoded text. The
transformation is specified by a secret key $K$ which is shared by the
sender and the receiver (see fig.~\ref{fig:crytoStructure})
This transformation is usually considered in a purely mathematical
framework, with no reference to any physical process. Yet,
Shannon~\cite{shannon:48,shannon:49} describes the generic steps of a
cryptosystem in terms of the repetition of diffusion and confusion
operations (see fig.~\ref{fig:crytoStructure}. Diffusion is a well
known physical process whose microscopical origin may be associate
with random walk.  It seems however that the contribution of physics
to classical cryptography has been only to provide some vocabulary but
no design principles.  The few physical devices that have been
proposed to encode a message are usually rather exotic and their
security is hard to certify~\cite{pourLaScience}

It should be noted that here we refer to cryptosystems using classical
transmission methods, i.e. bits of information traveling along an
electrical line.  We exclude the contribution of quantum physics for
which security does not rely on the difficulty of transforming
$M'$ back into $M$ but on the impossibility to intercept the information.

Thus, except from quantum mechanics, the contribution of physics to
cryptography has been surprisingly little. However, the second
principle of thermodynamics gives some interesting insights: the laws
of physics are such that in a closed system (e.g. a gas in a
container), time evolution always produces an increase of entropy
(i.e. decrease of information). In other words, all configurations
evolve to a similar final state, in which there seems to be no more
memory of the initial situation.  

This process is therefore a good encryption mechanism: the final
configuration reveals nothing about its origin. But what about
deciphering~?  Clearly, if there is no way to go back to the original
configuration, our methodology has little interest. It is a well known
paradox of classical physics that time always evolves to the future
and never to the past, despite the fact that the microscopic laws of
physics are fully symmetrical with respect to past and future.  There
should therefore be a way to come back. It is however highly
impracticable. Indeed, according to Newton's laws of mechanics, one
would have to reach every single particles of the system and to
exactly reverse their microscopic velocity. How can we ever do such a
thing in a real physical system~?  There is no way to act on each
particle and there is no way to know with full accuracy its current
velocity, not to mention that the smallest possible error we could
make in this reversing process will prevent the system to evolve back
to its initial state. Therefore, errors are an efficient way to prevent 
message decryption. Provided that we could control them, errors
thus offer a way to implement a secrete key.

Fortunately, we can export the above ideas in a framework where both
macroscopical irreversibility and microscopical reversibility
co-exist. This is the field of {\em discrete physics} exemplified by
the so-called Lattice-Gas-Automata (LGA) approach described in
section~\ref{section:LGA}. In a discrete physical world, one can act
on each particle with full accuracy, allowing us to run the system
forward and backward in time and to add well controlled errors.

This paper is organized as follows: we first define the LGA paradigm
and illustrate its behavior with respect to the second principle of
thermodynamics. Then we propose a complete cryptographic algorithm
based of the LGA approach in which Physics naturally suggests simple
and efficient operations, as well as a criteria to generalize the
algorithm to a wide range of topologies and sizes.  We provide some
evidence of the security (evolution of Hamming distance, flatness of
XOR table, computational effort to break the key with differential
cryptography). We finally  conclude by insisting on the deep and
promising link that exist between physics and cryptography.

\section{Discrete physics and Lattice-gas Automata}\label{section:LGA}

\begin{figure}
\centerline{\psfig{figure=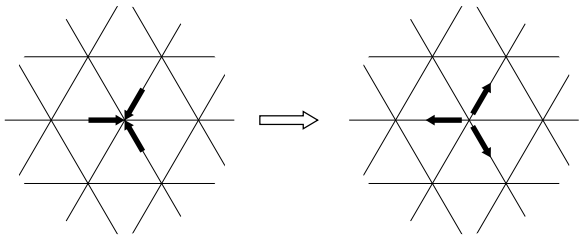,width=.6\textwidth}}
\caption{An example of a LGA, the so-called FHP model~\cite{fhp-prl}
defined on a hexagonal lattice where each site has  $z=6$ neighbors.}
\label{fig:fhp}
\end{figure}

Lattice-gas automata (LGA) are a special class of cellular automata
(CA) designed to provide a mesoscopic model of a physical system, such
as a gas or a fluid~\cite{BC-livre}. LGA can be thought of as a
virtual universe implementing a fully discrete abstraction of the real
world. Technically speaking, these automata consist of a $N$ point-particles
moving on a regular lattice in $d$ spatial dimensions, according to a
discrete time steps. The possible velocities of each particle are
restricted by the lattice topology in the sense that, in one time step
$\dt$, a particle move from one site to one of its existing
neighbor. Thus, if $z$ is the lattice coordination number, particles may have
$z$ possible velocities denoted $\v_i$, $i=1,...,z$.
Figure~\ref{fig:fhp} illustrates the situation.

Let us denote by $n_i(\r,t)$ the number of particles entering lattice
site $\r$ at time $t$ with velocity $\v_i$. Assuming that these
particles are not deflected at site $\r$, they will, at the next time
step, enter the neighboring site pointed by $\v_i$, with the same
velocity. In other words, $n_i(\r+\v_i\dt,t+\dt)=n_i(\r,t)$.  This
operation is called propagation and can be globally described by an
operator $P$. If $M(t)$ is a configuration of particles over the full
lattice, then 
\[ M(t+\dt)=P M (t) \]
represents the motion of all particles at every sites. Note that if
the lattice is not periodic in all direction, a special treatment is
required for the particles reaching the spatial boundaries.

We shall now assume that $n_i$ can be either 0 or 1 (no particle, or
at most one with velocity $\v_i$ at site $\r$ and time $t$). Also, we
assume that the particles entering the same site at the same time from
different direction (i.e. different velocities) interact according to
a pre-defined collision rule. The result of this collision is to
create new particles in some directions and to remove some particles
in other directions, as illustrated in fig.~\ref{fig:fhp}.

The collision process can be described by functions $C_i(n_1,
n_2,...,n_z)$, $i=1,...,z$,  taking the value 1 if the interaction produces a
particle with velocity $\v_i$ and 0 otherwise. After the collision,
the particles move to one of their neighboring site, according to the velocity
they have. Thus
\begin{equation}
n_i(\r+\dt\v_i,t+\dt)=C_i(n_1(\r,t), n_2(\r,t),...,n_z(\r,t))
\label{eq:lga}
\end{equation}
The fact that $C_i$ can be either 0 or 1 guarantees that the
occupation numbers $n_i$ obey our hypothesis of being 0 or 1 at all
times and all lattice positions. Therefore, the full dynamics can be
described by a set of $z$ bits of information at each lattice site and
each time steps.

As before, we can consider that $C$ is an operator which acts on all
lattice sites. Thus if $M(t)$ is the configuration at time $t$,
eq.~\ref{eq:lga} becomes
\[ M(t+\dt)=PCM(t) \]
where $P$ is the propagation operator.

Now, if we want our Boolean particles to behave macroscopically as
observed in a real physical system, the collision operator $C$ must be
carefully chosen. Its expression will crucially depends on the nature
of the physical process. For instance, it can be shown that, in the
appropriate macroscopic limit, fluid motion can be reproduced by such
a LGA. This is the case of the famous FHP model (2d) or FCHC models
(3d) (see~\cite{BC-livre}) which can be shown to
obey the Navier-Stokes equation of hydrodynamics. The proof of this
equivalence is mathematically rather
involved~\cite{BC-livre}. Intuitively, the LGA is a caricature of the
microscopic level but, at a macroscopic level, it behaves as a real
system. Over the past decade, this property has been intensively
exploited by many researchers to develop a new generation of
hydrodynamics solvers, such as  the so-called Lattice Boltzmann
method which keeps gaining in popularity to model and simulate complex
fluids.

When using this physical framework to build a cryptographic system one
property of direct importance is time reversibility. For a classical
physical system of particle, e.g. a fluid, it is well known that, by
modifying the velocity of each particle from $v$ to $-v$, but keeping
the same evolution equation, the full system retrace its own past. For
several reason, this time-reversal operation is out of reach in real
systems and also in standard molecular dynamics simulations.  First,
one needs to access all the degrees of freedom and, second, one must
know to infinite precision the actual value of the velocity. Finally,
in the case of a computer model, the calculation should not produce
the slightest numerical error.

\begin{figure}
\centerline{\psfig{figure=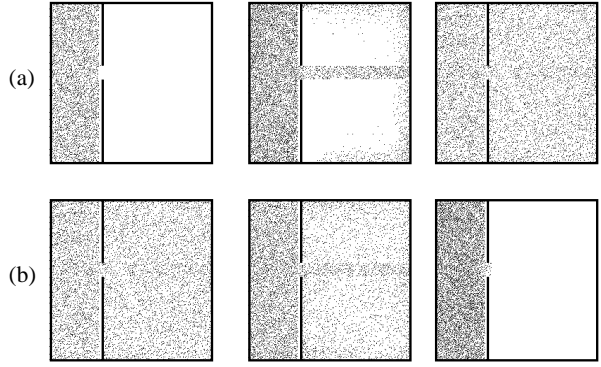,width=.6\textwidth}}
\caption{Time-reversibility in the so-called HPP LGA model. In (a),
  all particles are initially in the left compartment. Then they start
  expanding into the full space until an homogeneous situation is
  reached. Then, the velocity of all particles are inverted, as shown
  in (b) and the particle naturally trace back their way to the left
  compartment.}
\label{fig:reverse}
\end{figure}

However, in the case of a LGA model, reversing time is possible
because both the particles and the dynamics are Boolean. Calculations
are done with full accuracy, without truncations or errors. We
illustrate this behavior in fig.~\ref{fig:reverse}. When no errors are
introduced, it is possible to reverse the velocity of each particle by
applying, at a chosen time $t$, the reverse operator $R$ at each
lattice site and we see the system return to its initial state after $t$
repetition of propagation $P$ and collision $C$.

The operator $R$ permutes the value of the $n_i$ at each site so as to
place a particle of speed $\v_i$ the $-\v_i$ direction. Therefore
\begin{equation}
R^2=1
\label{eq:r2}
\end{equation}
For instance, in the case of a lattice with four directions
(North, East, South, and West), $R$ would swap East with West and
North with South. In this case
\begin{equation}
R(n_1,n_2,n_3,n_4)=(n_3,n_4,n_1,n_2)
\label{eq:reverse}
\end{equation}
where $\v_1=-\v_3\quad \v_2=-\v_4$.
An interesting
relation between $P$ and $R$ is
\begin{equation}
PRP=R
\label{eq:prp}
\end{equation}
which means that reversing allows the particle to propagate in the
direction they came from and reach it with opposite velocity (which is
the reason why $PRP=R$ and not $PRP=1$).  

The reversibility of the
collision operator means
\begin{equation}
CRC=R
\label{eq:crc}
\end{equation}
because reversing the post collision configuration and applying the
collision again give the pre-collision state, but with opposite
velocities.
 
With $PRP=R$ and $CRC=R$ one has
\begin{eqnarray}
(CP)^rR(PC)^r &=& (CP)^{r-1}CPRPC (PC)^{r-1} \nonumber \\
              &=&(CP)^{r-1}R (PC)^{r-1}=...R \nonumber
\label{eq:reverse_simple}
\end{eqnarray}
which shows that performing $r$ iterations, reversing the velocity and
performing again $r$ iterations take us back to the initial
configuration, with all particles having an opposite velocity. This proves
the time-reversibility of the process.

\begin{figure}
\centerline{\psfig{figure=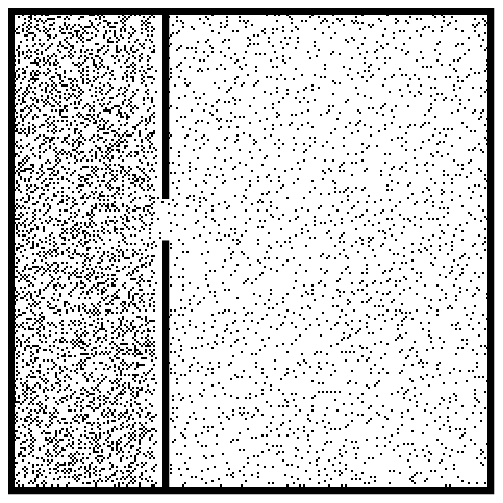,width=.4\textwidth}}
\caption{Effect of one error on the time-reversibility of the HPP LGA
  model. Compared with the situation of~\ref{fig:reverse} (b), one
  extra particle is added on the homogeneous state. As a result, the
  particles are no longer able to return to their initial
  position, in the left compartment. }
\label{fig:reverse-error}
\end{figure}

The time-reversal operation is possible because the system has, in
fact, not forgotten about its initial condition (in
fig.~\ref{fig:reverse} all particles are in the left compartment)
even though it evolves to a state where this information seems
completely lost. The information is actually not lost but diluted on
all degrees of freedom. Therefore, if one perturbs just a single piece
of this distributed information, the system loses its capability to
evolve back to its past. This is shown in fig.~\ref{fig:reverse-error} where
a particle has been artificially added before reversing the
time. Clearly, all the particles whose trajectory have been directly
or indirectly affected by the presence of this extra particle are
unable to return to their original position. This single modification
creates an avalanche of perturbations which expands further and
further as the number of time steps increased.
This experiment illustrates how in a LGA paradigm we can play with the
second principle of thermodynamics and reverse the arrow of time. At
the macroscopic level, an H-theorem can be proved (provided that $C$
is properly chosen), meaning the an entropy can be defined and that it
always ends up growing. At the microscopic level, however, past and future
are fully equivalent provided one has full knowledge of all degrees of
freedom of the system.

Another point worth mentioning about LGA is its intrinsic parallelism. All the
particles move synchronously on the lattice and can be updated at the
same time. The large degree of parallelism present in physical systems
is naturally reproduced here and can be exploited to speedup
dramatically the simulation.  The collision operation is 
embarrassingly parallel whereas the propagation requires regular
collective communications.

\section{A new cryptography algorithm}\label{section:algo}

In this section we make explicit the link we claimed to exist between
classical symmetric cryptographic algorithms and a LGA-fluid. Then we
propose a new cryptographic algorithm which exploits this
physical analogy and  we discuss the properties of such an  approach.

\subsection{Standard symmetric block-cipher}\label{section:standard}

Symmetric cryptography (as opposed to public-key cryptography) assumes
that prior to the exchange of the encoded message, the sender and the
recipient have agreed on a secret key which is used, on one side to
encode the plain text message and, on the other side to decode it.

Block-ciphers (as opposed to stream-ciphers) consider that the
information that need to be transmitted can be divided into blocks of,
say, $N$ bits. Each block is encoded by the algorithm and sent before
the next block is processed.

Symmetric cryptosystems are usually described as a combination of two
operations, termed diffusion and confusion~\cite{shannon:48,shannon:49} These
operations are repeated in alternance and iteratively on the message
$M$ to be encoded. Each application of the diffusion and confusion
step is called a round. Thus the full process consists in applying $r$
rounds on message $M$. The larger the value of $r$, the more secure
the encoding but the slower the encryption process.

In practice, confusion is achieved by transforming each byte $b$ of
$M$ according to a {\emph substitution box}, or $S$-box, which is a
given invertible, non-linear function. Diffusion corresponds to a
deterministic shuffling of the bytes across the full message.

Following Kirckhof principle, the secrecy of the process is not
obtained by keeping hidden the algorithm but by parameterizing some of
its aspects by a secret key $K$, made up of $N'$ bits only known to
the communicating persons.  A common strategy is to 
XOR the key bits with the message bits before each round. In order to
improve the security of the encoding process, the bits of the key are
not reused identically over all rounds.  They are transformed after
each round, for instance by applying the same diffusion-confusion
process as used for the message itself. These new sets of bits
obtained at each round are called {\emph round keys}.

Note that, in some ciphers, the secret key $K$ is  used to
parametrized the $S$-box rather than being a secret, dynamic mask.

Quite often, the key size $N'$ matches the block size $N$. When the
key is smaller than the message, padding is necessary. This can be a
security issue if these additional bit are not carefully chosen.  When
the key is larger than the message, the extra key bits can be used to
produce the successive round keys. When the entire message is composed
of several blocks, the final round key of block $i$ is usually used as
the first round key for block $i+1$. Once the recipient has received
all blocks, decoding can start by inverting the ciphering process. For
this purpose, the recipient must compute the final round key.

The above general discussion applies to well known cryptographic
algorithms such as DES. AES, IDEA~\cite{schneier:96,daemen:02}

\subsection{A physically based cryptography algorithm}

By comparing the material of sections~\ref{section:LGA} and
\ref{section:standard}, we clearly see how the LGA-fluid bears
structural similarities with the classical block-cipher engines.

The plain text message (block) corresponds to the initial
configuration of the physical particles (one bit is the presence or
absence of a particle), the $S$-box is the collision process and
diffusion is obtained by our propagation. The $r$ rounds represent a
time evolution (iterations) and decoding reflects time-reversal
invariance. The role of the key is a bit less obvious but it can be
thought of as a controlled error which is added to break the possibility
to reverse time. Somehow, the round keys reproduce the space-time
evolution of a secret configuration of particles that interact with
the ``message'' particles.

However, some specificities of the physical process need to be
emphasized with respect to the classical cryptographic algorithms.  

\paragraph{Space dimension}
Our particles move in a two-dimensional space. The dynamics can be
generalized to higher dimensions. Usually, in cryptography the concept
of space is absent, or little exploited (AES uses a matrix formulation
which is more a mathematical construct than the desire to embed the
process in the physical space). In physics, it is known that higher
dimensional spaces allows for more efficient mixing mechanisms and
reduces the space diameter.

\paragraph{Fine grain diffusion}
In the LGA model, particles are mixed across space through the
propagation process. Its purpose is to dilute the information
carried by the initial configuration on all degrees of freedom.  This
``diffusion''\footnote{In physics, diffusion refers to a random walk,
  which is not what our propagation does} takes  place at a  bit
level, whereas it is usually made a byte level in standard cryptography.

\paragraph{Collision}
Like a $S$-box, collision is a local operation since it is repeated
independently over all lattice sites. It can be implemented as a
lookup table and acts upon $z$ bits where $z$ is the lattice
coordination number. Thus, depending on the lattice topology,
collision  acts on pieces of information that can be smaller, larger
or equal to a byte. In the discrete-fluid, $C$ is
reversible, that is the same function is used forward in time or
backward in time. This is a significant simplification when hardware
implementation is considered. On the other hand, in a physical model,
the collision $C$ may have some undesired properties with respect to
cryptography. In our fluid example, the number of particles is
conserved by $C$ (because Nature does). These conservations laws must
be removed when devising a cryptographic scheme.

\paragraph{Scalability and parallelism}
The LGA fluid is composed of the juxtaposition of many identical
units: the lattice cell. In addition, these lattice cells are locally
connected according to a simple topology. Therefore, system of
arbitrary size can be constructed with the same design principle and
same simplicity. The locality of the collision and the regularity of
the diffusion allows for massive parallelism (one processing element at
each lattice site, for any lattice size).

The above discussion shows that classical mechanics, in the flavor of
a discrete physical system (LGA or CA model), offers a natural
framework for designing a cryptographic device. It actually resembles
the classical designs obtained from  mathematical principles but
has some interesting features: cipher of any size can be produced with
the same design principle, in which parallelism is ensured. Moreover,
simplicity, regularity and adequation of the 2D physical space with
the spatial layout of integrated circuits allows fast hardware
implementation.

\subsection{A specific instance of the algorithm}

\begin{figure}
\centerline{\psfig{figure=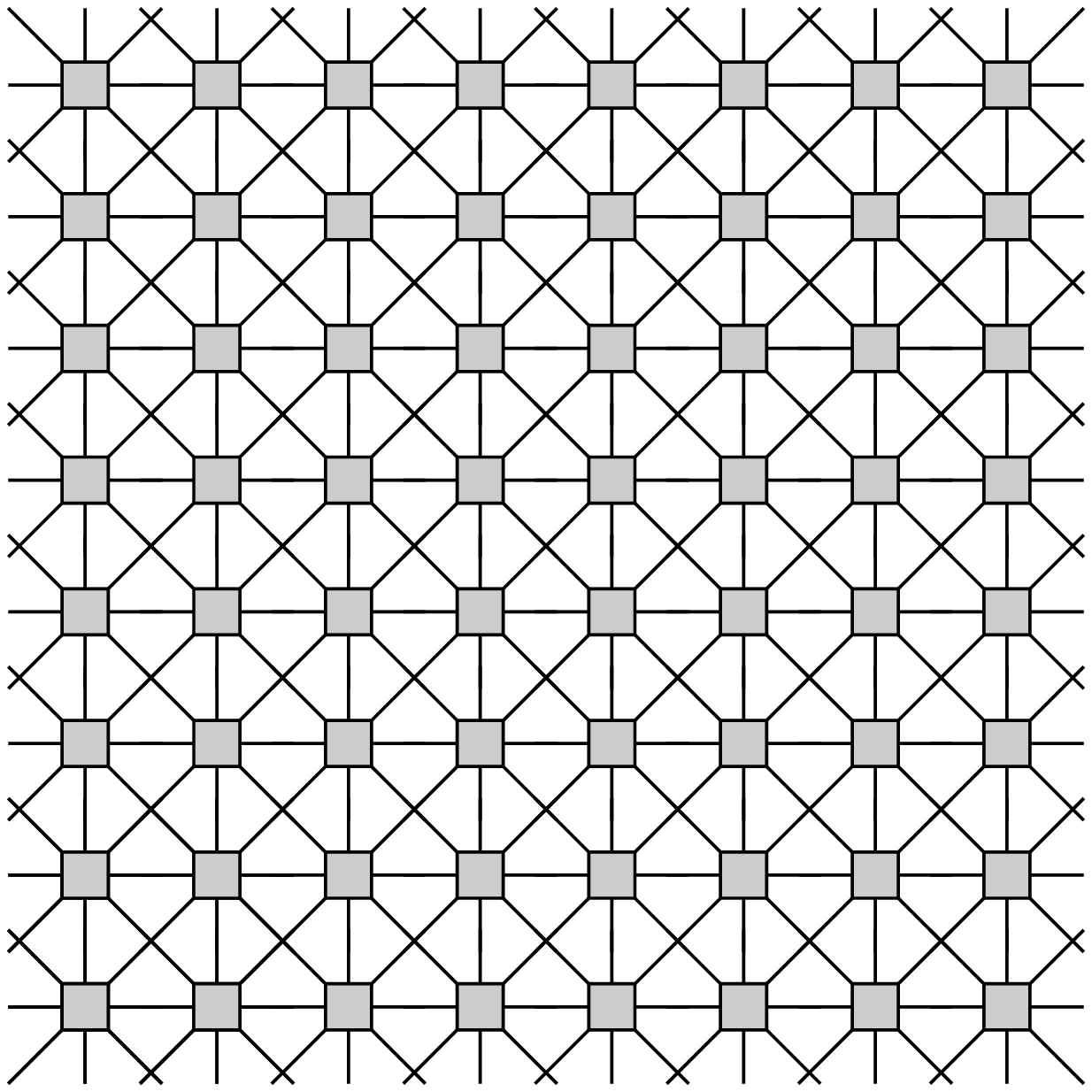,width=.6\textwidth}}
\caption{The 2D lattice used in the \Crypto algorithm. Here, a system
  of size $8\times8$ is shown. Each gray box contain 8 bits so that
  the full block has 512 bits. Here we assume a periodic topology:
  left and right borders are wrap around, as well as the upper and
  lower ones.}
\label{fig:d2q8}
\end{figure}

We can now define more precisely an instance of a cryptographic
algorithm based on the physics analogy. We consider a 2D square
lattice, with $z=8$ links (up, left, right, down, plus the four
diagonals) as illustrated in fig.~\ref{fig:d2q8}. In the LGA jargon,
this topology is often referred to as D2Q8 (2 Dimensions, 8
Quantities).
The lattice is a square of size $\sqrt{N/z}\times
\sqrt{N/z}$, with periodic boundaries. Each site contains $z=8$ bits
(i.e. a byte) and the full lattice can encrypt a block $M$ of $N$ bits. The
propagation $P$ and reverse operator $R$ are as described in
section~\ref{section:LGA}: each of the eight bits of a given site is
moved to one of the eight neighbors of that site and $R$ swaps the
bits traveling in opposite direction.

In order to build a collision which is reversible (i.e $CRC=R$), but
has no other symmetries, we first build a random, self-inverse
function $C'$ such that $C'C'=1$. Then, clearly,
$C=C'R$ is reversible because
\[ CRC=C'RRC'R=C'C'R=R \]
To produce $C'$ one has to define the image $j=C'(i)$ for all
$i=0,1,...2^z-1$. We start with $i=0$ and choose $j$ at random. Then
we immediately set $C'(j)=i$ to ensure idempotency. We then proceeds
similarly with the next $i$ for which the image or pre-image has not
yet been computed. The operator $C'$ obtained in this way can be further
tested if needed to ensure that has no accidental undesired features.
The function $C=C'R$ is then published as part of the algorithm

The secret key $K$ has $N'$ bits. We assume that $N'\le N$. If $N'<N$
some padding is needed. We produce the $N-N'$ remaining bits by
applying for instance the Kirkpatrick-Stoll
procedure~\cite{kirkpatrick:81} to build random bit with density 1/2 out
of an initial set: $b_\ell=b_{\ell-250}\oplus b_{\ell-103}$ which can
always be applied if $N'\ge 250$. Note that lagged Fibonacci
method~\cite{marsaglia:85} can also be used: $b_\ell=b_{\ell-55}\oplus
b_{\ell-24}$.

Then the round
key at iteration $m$ is obtained by successive application of $PC$
over the previous round key at iteration $m-1$. Finally the round keys
are combined with the iterated message with an {\tt XOR}. 

In order to ensure that the encryption and decryption are completely
identical, it is convenient to start the process by a reverse and a
propagation steps. This gives the following algorithm (whose structure
remain the same even with a different topology)
\begin{verbatim}
algorithm Crystal(M,K)
  reverse(M), reverse(K)
  propagation(M), propagation(K)
  repeat r times
    M=M xor K
    collision(M),  collision(K)
    propagation(M), propagation(K)
  end repeat
  M=M xor K
  return M, K
end algorithm
\end{verbatim}
where {\tt XOR} is the same as $\oplus$  and $K$ refers to the padded
key.  In appendix A, it is demonstrated that this scheme is
reversible, namely that if $(M',K')=\Crypto(M,K)$ then $(M,K)=\Crypto(M',K')$.

The appropriate value of the number of round $r$ is discussed later.

\section{Properties}

In this section we derive some important properties of our
cryptographic algorithm. The results will show that we can encrypt
large blocks of data by taking a large enough lattice. We will
show that this process increases both throughput and security.

\subsection{Hamming distance}
First we consider the number $r$ of iterations needed to make two
initial messages $M_1$ and $M_2$ as different as possible from each
other. We assume that messages $M_1$ and $M_2$ are identical
except for one bit. The speed at which these two initial conditions
diverge from each other reflects the discrete Lyapounov exponent of the
dynamics. The question is to determine how many steps are necessary so
that the single bit error  has ``polluted'' the full system. From that
point on, all degrees of freedom have been informed that the two
message actually differ.

Since the key evolves similarly in the two messages, it is irrelevant
and can be omitted from the discussion. After each
collision-propagation step, information propagates away from the
initial source of error. For the D2Q8 topology we discuss here, it is
easy to show that the number of sites that can be reached from an
initial lattice position grows with the number of round $r$ as a
square shaped region $A(r)$ of side $2r+1$ (and  of area
$(2r+1)^2$). This relation simply reflects that, at round $r+1$, all
lattice sites that are bordering $A(r)$ according to the D2Q8 topology
will belong to $A(r+1)$.

Thus, when $(2r+1)^2=N/z$, all the sites are informed. This value of
$r=(1/2)\sqrt{N/z}$ corresponds to the {\em diameter} $d$ of the
lattice.

\begin{figure}
\begin{tabular}{m{.5\textwidth}m{.5\textwidth}}
  \psfig{figure=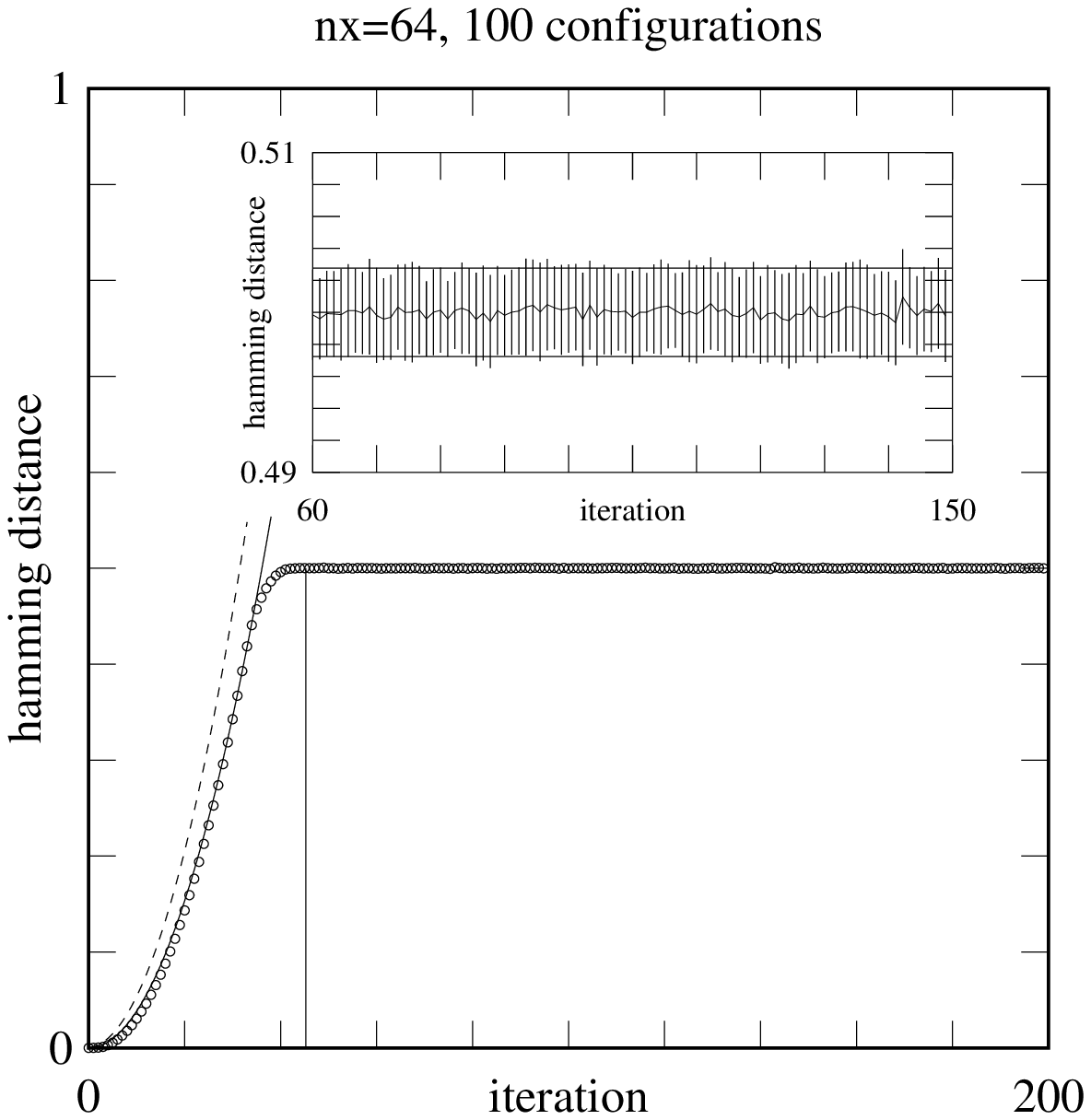,width=.5\textwidth} &
  \psfig{figure=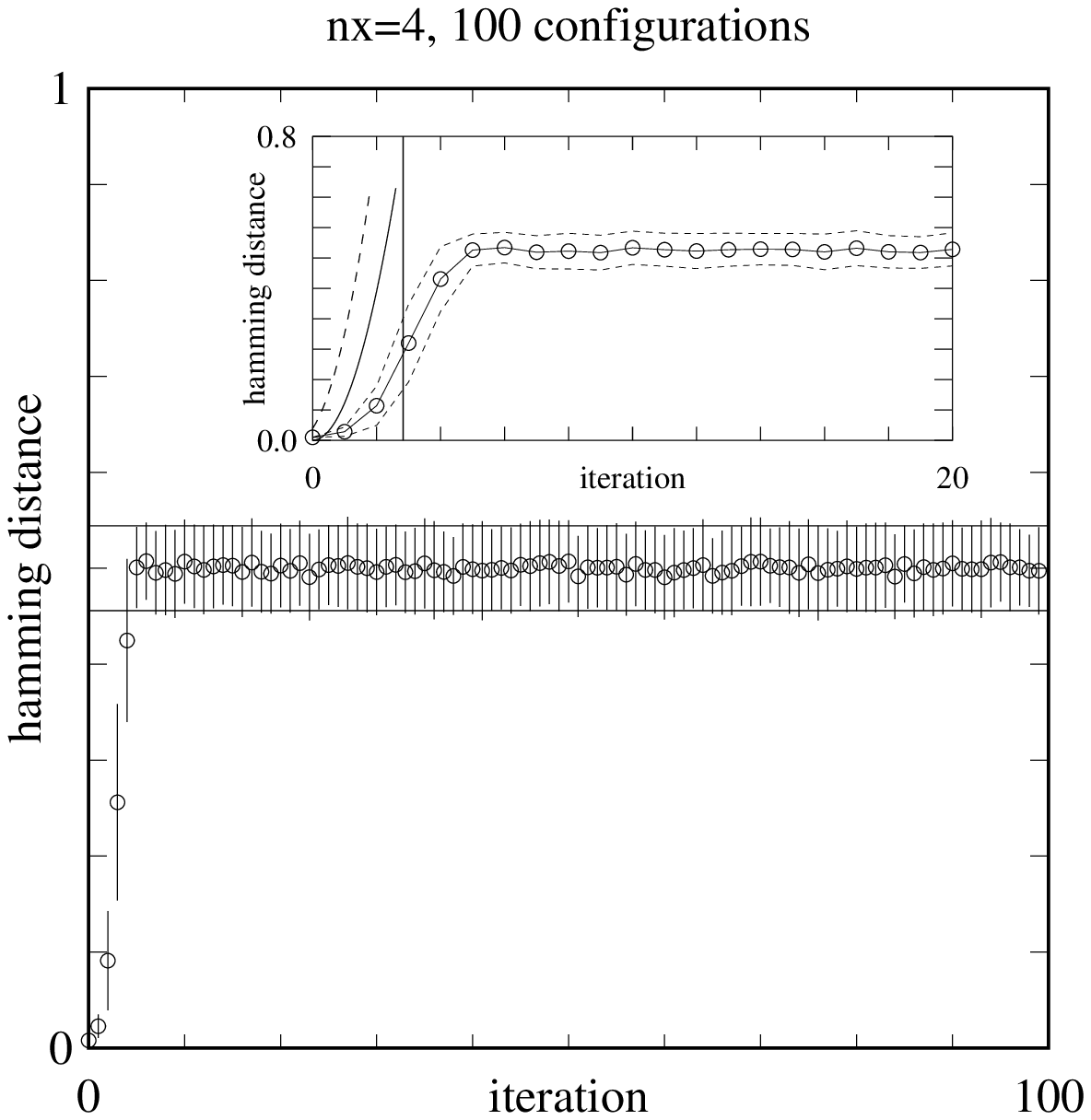,width=.5\textwidth} \\
\centerline{(a)} & \centerline{(b)}
\end{tabular}
\caption{Evolution of the Hamming distance between two messages
 initially differing only by one bit. In (a) we have
 $N=64\times64\times8=32768$ bits and in (b) $N=4\times4\time8=128$ bits.
 Comparison with the ideal curve (eq.~\ref{eq:hamming-best} is given
 with the doted parabola. The solid line parabola is the theoretical
 estimate of eq.~\ref{eq:speed}). Finally, the vertical line show the
 iteration at which, according to eq.~\ref{eq:plateau}, the plateau
 should be reached.}
\label{fig:hamming}
\end{figure}

\begin{figure}
\begin{tabular}{b{.5\textwidth}b{.5\textwidth}}
\centerline{\psfig{figure=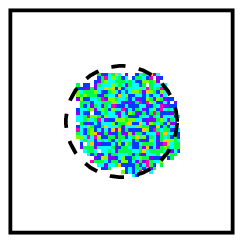,width=.4\textwidth}}&
\centerline{\psfig{figure=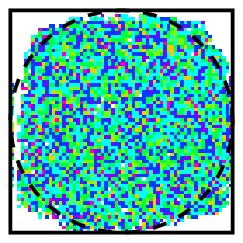,width=.4\textwidth}}
\end{tabular}
\caption{Snapshot of the error propagation region, after 16 and 32
  iterations, in a system of size $64\times64$. The non-blank regions 
indicates where the two configurations differ. The darker the gray,
  the more are the bits that differ. The dashed-line disks
  have radius 16 and 32, respectively; thus, the error propagates at
  speed one for this topology.}
\label{fig:disk}
\end{figure}

Since there are $z$ bits per sites and two random patterns differ on
average by half of their bits, the Hamming distance between $M_1$ and
$M_2$ after $r$ steps is expected to ideally evolve as
\begin{equation}
H={1\over2}z(2r+1)^2
\label{eq:hamming-best}
\end{equation}
Figure~\ref{fig:hamming} shows the evolution of the normalized Hamming
distance $h(r,M_1,M_2)=H/N$ in the case of our algorithm. We observe
that, essentially, a number $r=\sqrt{2} d$ is needed to reach the
plateau $h=1/2$ where each bit of the two configurations differ
randomly. Thus, fluctuations of magnitude $1/\sqrt{N}$ around the
value 1/2 are expected.
The speed at which the $h=1/2$ plateau is reached is less than
predicted by eq.~\ref{eq:hamming-best} because after a collision (or
substitution), only about  $z/2$ bits differ from the reference configuration.
As shown in  fig.~\ref{fig:disk}, the error
thus propagates as a disk and not a square. The diameter of that error-disk 
grows on average by one lattice site at each
iteration. Thus, during the first $r=\sqrt{N/z}/2$, $H$ behaves as
\begin{equation}
H={z\over2} \pi r^2 
\label{eq:speed}
\end{equation}
When the error disk has reach the boundary of the lattice, the Hamming
distance grows slower. A few more steps are needed to reach the corner
of the domain. Assuming that the radius of the disk keep growing at
speed 1, the total number of iterations is equal to half the length of
the diagonal, i.e.
\begin{equation}
r={\sqrt{2}\over 2}\sqrt{ {N\over z} }
\label{eq:plateau}
\end{equation}
This is the minimal number of rounds  needed to mix the
information all over the system. We will argue below that more rounds
may be needed to ensure more security against cryptanalysis. Thus we
write $r=\alpha d$ where $\alpha$ is some constant. In the
case of a D2Q8 lattice, we obtain
\begin{equation}
r=\alpha \sqrt{N/z}
\label{eq:min_diameter}
\end{equation}

\subsection{Throughput and security}

Let us assume that the number of rounds is $r=\alpha \sqrt{N/z}$, as
discussed in the previous section. If the process is fully
parallelized, propagation and collision take a constant time,
$t_{pc}$. Then the time $T$ needed to encrypt the $N$ bits of the
message is
\begin{equation}
T=t_{pc} r= t_{pc} \alpha \sqrt{N/z}
\label{eq:time}
\end{equation}
Therefore, the encryption throughput $W$ is 
\begin{equation}
W={N\over T}= { \sqrt{z} \over \alpha t_{pc}} \sqrt{N}
\label{eq:throughput}
\end{equation}
Thus, when large data blocks are encrypted and full
parallelism is implemented, the throughput increases, even though the
number of round also increases. However, the number of round increases slower
than the data size.  

On the other hand, when the same $N$ bits are split in several smaller
blocks (as is the case with standard cryptosystems), the total
throughput is that obtained with one block and does not increase. On
large enough data set, our approach, with full parallelism, will
always give a better performance. Estimate shows that for $N=2048$, a
throughput of 10 Gb/s is expected with a FPGA implementation. Note
that, the hardware simplicity of our cipher makes it possible to
obtain high throughput even on small data sets.

Security is also improved by our approach. Indeed, it is well admitted
that the main factor impacting  security in a symmetric block cipher is the
number of rounds. In our case, the number of round increases as $N$
grows, making the cipher more resistant to cryptanalytic attacks. Thus
security is improved at the same time as throughput increases.

In the next section we give an estimate of the difficulty to break our
algorithm when differential cryptanalysis  is used.

\section{Differential cryptanalysis}

\subsection{Introduction}

As already mentioned, a cryptosystem is a recipe to transform of a
plain text message $M$ of $N$ bits into another message $M'$, also
containing $N$ bits. Since the process must be invertible in order to
achieve decryption, the transformation must be a {\emph permutation}
from the set ${\cal M}_{_N}$ of all $N$-bit messages into itself. The
secret key $K$ can then be seen as parameter which selects one of the
possible permutation.

With $N$ bits, there are $2^N$ possible messages and $(2^N)!$ possible
permutations. This is by far too large a number to use all possible
permutations. That would require very long keys (with
$\log_2(2^N!)\approx (N-1)2^N$ bits) to index all of them.  

Therefore, in practice, a cryptosystem is a {\emph subset} of the
possible permutations from ${\cal M}_{_N}\to{\cal M}_{_N}$, with a well
specified indexing scheme to properly associate each key with each
accessible permutation. Typically, $K$ has also $N$ bits and can index
$2^N$ permutations only. Obviously the relation between the key and the
permutation is not explicit and is hidden by the algorithm which
tells how to compute $M'$ from $M$ and $K$.

The task of cryptanalysis is to obtain information on the secret key
$K$ from the knowledge of some pairs $(M,M')$. In the case where all
$(2^N)!$ permutations are possible, the knowledge of $(M,M')$ gives
little information on the chosen permutation, hence on the value of
$K$. Indeed, if $(M,M')$ belongs to the secret permutation, there are
still $(2^N-1)!$ permutations to search for. 

The other extreme is
given by the following very simple cryptosystem: $M'=M\oplus K$. This
is a rather trivial indexing scheme since $K$ is immediately determined
by $(M,M')$ as $K=M\oplus M'$.

From a combinatorial point of view, the situation is not very
different with practical cryptosystems. With $2^N$ possible
permutations of a set of $2^N$ messages, a pair $(M,M')$ is likely to
belongs to only one permutation. Thus, in principle, $(M,M')$ contains a
lot of information on $K$. However, this information is usually quite
difficult to obtain explicitly because the relation $K=K(M,M')$ is
expected to be quite intricate.

The goal of differential cryptanalysis is to obtain information on the
key $K$ by considering how two plain text messages $M_1$ and $M_2$ get
encrypted into $M_1'$ and $M_2'$. Let us know consider such an attack
in the case of our cryptosystem. 

\subsection{Complexity to break the \Crypto algorithm}

With $M_i^{(m)}$ and $K^{(m)}$ denoting the state of the messages and the
key after $m$ rounds, the algorithm \Crypto gives
\begin{equation}
M_i^{(m)}=PC\left(M_i^{(m-1)} \oplus K^{(m-1)}\right)
\end{equation}
for $i=1,2$. By XORing the above relation for $i=1$ and $i=2$ and
applying inverse propagation, we obtain
\begin{equation}
P^{-1}\left(M_1^{(m)}\oplus M_2^{(m)} \right)=
  C\left(M_1^{(m-1)} \oplus K^{(m-1)} \right)\oplus
  C\left(M_2^{(m-1)} \oplus K^{(m-1)} \right)
\label{eq:c-diff1}
\end{equation}
It is now convenient to introduced $\Finv$, an inverse XOR function
associated with $C$. Suppose that
\[ b=C(a_1)\oplus C(a_2) \]
for some known $z$-bit value $b$ and unknown $z$-bit values $a_1$ and
 $a_2$. The question is to know which pairs $a_1$ and $a_2$ can
 possibly produce the given $b$. More specifically and for reasons
 that will become clear in a moment, we want to know the value of
 $a_1\oplus a_2$ compatible with $b$. So we write
\begin{equation}
a_1\oplus a_2 \in \Finv(b) \qquad{\rm iff}\qquad b=C(a_1)\oplus C(a_2)
\label{eq:Finv}
\end{equation}
For a given collision operator $C$, $\Finv$ can be computed by
exhaustive search. In practice one has to generate all possible values
of $a=a_1\oplus a_2$ and compute what $b=C(a_1)\oplus C(a_2)$ it
produces. Since we consider $z$-bit values, there are $2^z$ ways to
choose $a$ and $2^z$ possible values for $b$. Therefore, $\Finv$ can
be represented as a $2^z\times 2^z$ matrix. A zero element at position
$(a,b)$ in this matrix means that the corresponding $b$ cannot be
produced with the corresponding $a$. Non-zero elements are set to the
number of pairs $(a_1,a_2)$ such that $C(a_1)\oplus C(a_2)=b$ and
$a_1\oplus a_2=a$.  Indeed,  a given $b$ can be obtained several times with
the same $a$, since each $a$ can be produced from $2^z$ combinations
$a_1\oplus a_2$, where $a_1$ is free and $a_2=a\oplus a_1$. Therefore,
the sum of each  row of the matrix is $2^z$.

Similarly, each column also sums up to $2^z$. To see it, assume that
$b$ is written as $b_1\oplus b_2$ where $b_2=b\oplus b_1$. There are 
$2^z$ such values of $b_1$. As $C$ is invertible, we can compute
$a_1=C^{-1}(b_1)$ and $a_2=C^{-1}(b_2)$. This pair $(a_1, a_2)$ will
satisfy that $C(a_1)\oplus C(a_2)=b$. Thus a given $b$ is obtained
$2^z$ times, distributed over the values of $a=a_1\oplus a_2$.

By normalizing each entry with $2^z$, one obtains a matrix whose rows and
columns sum up to 1. An example for $z=4$ and our procedure to build
a random  reversible $C$ is given below.
\goodbreak
{\tiny
\begin{tabular}{r|rrrrrrrrrrrrrrrr}
    & 0  & 1  & 2  & 3 &  4  & 5 &  6  & 7 &  8  & 9 & 10  &11 &  12
     &13 & 14  &15\\
\hline
 0 &  1&  0 &  0 &  0 & 0  & 0  & 0  & 0  & 0  & 0  & 0  & 0  & 0  & 0  & 0  & 0 \\
 1 &  0& 1/8& 1/8& 1/8&   0&   0&   0& 1/8& 1/8&   0&   0&   0& 1/4&     0&   0& 1/8\\
 2  &0   &0   &0  &1/4 &1/8 &  0 &1/8 &  0 &1/8 &  0 &  0 &1/8 &1/8 &1/8 &  0 &  0\\
 3 & 0  & 0  &1/8  & 0  &1/4   &0   &0  &1/8 & 1/8  &  0  &  0  &  0 & 1/8  &  0 & 1/8&  1/8\\
 4 & 0 & 1/4 & 0 &  0  &1/8 &  0 & 1/4 & 1/8 &  0 & 1/8 & 1/8 &  0 &  0 &  0 &  0 &  0\\
 5 & 0 &   0 & 0 &  0 &  0 & 1/4 & 1/4 &  0 &  0 & 1/8 & 1/8 &  0 &  0 & 1/8 & 1/8 &  0\\
 6 & 0 & 1/8& 3/8 &  0 &  0 & 1/8 &  0 & 1/8 &  0 &  0 & 1/8 & 1/8 &  0 &  0 &  0 &  0\\
 7 & 0 &   0 &1/8 & 1/8 &  0 & 1/8 & 1/8 & 0 & 1/8 &  0  &1/8 &  0 &  0 &  0 & 1/4 &  0\\
 8 & 0 &   0& 1/8 &  0 & 1/8 &  0 &   0 & 0 &  0 & 3/8 &  0 &  0 & 1/8 & 1/8 & 1/8 &  0\\
 9 & 0 & 1/4 & 0 &  0 &  0 & 1/4 &   0 & 0  &1/8 &  0 &  0 & 1/8 &  0 & 1/8 & 1/8 &  0\\
10 & 0 & 1/8 & 0 &  0 &  0 & 1/8 &   0 & 0 &  0 & 1/8 & 1/4 &  0 &  0 & 1/8 &  0 & 1/4\\
11 & 0 &   0& 1/8 & 1/8 &  0 & 1/8 & 1/8 & 0 &  0 &  0 &  0 &  0 & 1/8 & 1/4 & 1/8 &  0\\
12 & 0 &   0 & 0 & 1/8&  1/8 &  0 &   0 & 0 & 1/4 &  0 &  0 & 1/8 &  0 & 1/8 & 1/8 & 1/8\\
13 & 0 & 1/8 & 0 &  0 & 1/8 &  0 &   0& 1/4 &  0 & 1/8 &  0 &  0 & 1/8 &  0 &  0 & 1/4\\
14 & 0 &   0 & 0 & 1/8 &  0 &  0 & 1/8 & 0 & 1/8 & 1/8 &  0 & 3/8 &  0 &  0 &  0 & 1/8\\
15 & 0 &   0 & 0 & 1/8 & 1/8 &  0 &   0 &1/4 &  0 &  0 & 1/4 & 1/8 & 1/8 &  0 &  0 &  0\\
\end{tabular}
}
\goodbreak
For instance, the element $1/8$ obtained for $a=4$ and $b=7$ means
that if $b=7$, the probability that $a=4$ is $1/8$. Clearly, if $a=0$,
it means that $a_1=a_2$ and thus $b=C(a_1)\oplus C(a_2)=0$, and
conversely.

We can now come back to eq.~\ref{eq:c-diff1}. With
definition~\ref{eq:Finv}, we can rewrite it as
\begin{eqnarray}
\Finv  P^{-1}\left(M_1^{(m)}\oplus M_2^{(m)} \right) &=&
  M_1^{(m-1)} \oplus K^{(m-1)} \oplus
  M_2^{(m-1)} \oplus K^{(m-1)}  \nonumber \\
  &=&   M_1^{(m-1)} \oplus  M_2^{(m-1)}
\label{eq:c-diff2}
\end{eqnarray}
By repeating this relation recursively, one obtain
\begin{equation}
M_1^{(1)} \oplus  M_2^{(1)} = \left(\Finv  P^{-1}\right)^{r-1}\left(M_1^{(r)}\oplus M_2^{(r)} \right)
\label{eq:c-diff3}
\end{equation}
where $r$ is the number of rounds. Below we will show that if
$M_1^{(1)} \oplus M_2^{(1)}$ is known to the attacker, it is rather
easy to obtain the secret key $K$. The question we want to investigate
first is how much computational effort is required to obtain
$M_1^{(1)} \oplus M_2^{(1)}$ from $M_1^{(r)}\oplus M_2^{(r)}$ which,
by hypothesis, is known since attackers are supposed to have access to
any pairs $(M,M')$ they want. The estimate of the complexity of
finding $M_1^{(1)} \oplus M_2^{(1)}$ is given below.

Since we assume that $r>d$, where $d$ is the lattice diameter,
$M_1^{(r)}$ and $ M_2^{(r)}$ differ over all $N/z$ lattice sites. In
order to perform the backward scheme indicated in
eq.~\ref{eq:c-diff2}, one has to find all possible pre-images by
$\Finv$ of $P^{-1}\left(M_1^{(m)}\oplus M_2^{(m)}
\right)$. Empirically we observe that the number of pre-image of a
given $b$ is larger than $2^z/4$. Of course this depends on the choice
of $C$, but this seems to be a minimal value.  Therefore, for each
lattice site, at least $2^{z-2}$ values are possible for $M_1^{(r-1)}
\oplus M_2^{(r-1)}$. This requires to select $(N/z)2^{z-2}$ candidates
for $M_1^{(r-1)} \oplus M_2^{(r-1)}$.

The same argument can be repeated $r-d$ times. After that, we can
quickly exclude some possibilities. Indeed, at this point, we know
that the error has not been able to propagate up to the outer boundary
of the lattice. For these lattice sites, $M_1^{(d-1)} \oplus
M_2^{(d-1)}$ must be zero. Thus the number of sites for which the
exploration continues is $(\sqrt{N/z}-2)^2$. If we undo one more step,
even more possibilities can be excluded and the pre-images of ``only''
$(\sqrt{N/z}-4)^2$ sites must be investigated. Following this idea for
the $d-1$ steps, one has to explore $3^2\times 5^2\times...\times
(\sqrt{N/z}-1)^2$ possible configurations\footnote{for a D2Q8
topology}, each with $2^z/4=2^{(z-2)}$ possible values. 

An inferior bound for this number is (see appendix~\ref{appendix-2})
\[ \left(3^2\times 5^2\times...\times(\sqrt{N/z}-1)^2\right)2^{z-2}>
(d/2)^{2d}2^{z-2}= {1\over4}\left({N\over z}\right)^{d}2^{z-2} 
\]

Thus, in total (undoing the rounds beyond and below the diameter) implies to
investigate 
\begin{equation}
{\cal N}>(N/z)^{r-d} 2^{(z-2)(r-d)} (N/z)^{d} 2^{z-4}=
            (N/z)^{r} 2^{(z-2)(r-d)+(z-4)}
\label{eq:total-work}
\end{equation}
candidates for $M_1^{(1)} \oplus M_2^{(1)}$.

This relation will be discuss in detail in
section~\ref{section:perf-security}. Some values of ${\cal N}$ as a
function of $N$ and $r$ are given in table~\ref{table:1}.

\subsection{The last step to obtain the key}

In this section, we show how to compute the key from a possible candidate
$M_1^{(1)} \oplus M_2^{(1)}$. Since $M_i^{(1)}=PC(M_i^{(0)}\oplus
K^{(0)})$ we obtain
\begin{equation}
C\left(M_1^{(0)}\oplus K^{(0)}\right) \oplus 
C\left(M_2^{(0)}\oplus K^{(0)}\right)
=P^{-1}\left(M_1^{(1)}\oplus M_2^{(1)}\right)
\label{eq:c-diff4}
\end{equation}
Let us introduce $b_1=C\left(M_1^{(0)}\oplus K^{(0)}\right)$, 
$b_2=C\left(M_2^{(0)}\oplus K^{(0)}\right))$ and 
$b=P^{-1}\left(M_1^{(1)}\oplus M_2^{(1)}\right)$.
The quantity $b$ is supposed to be known. 

We now investigate all possible values of $b_1=0,1,...(2^z-1)$.
For each of these $b_1$, $b_2$ can be computed as
\[ b_2= b\oplus b_1 \]
Since $C$ is invertible, we can compute from the definition of $b_1$
\[ M_1^{(0)}\oplus K^{(0)}  = C^{-1}(b_1) \]
Thus, from the definition of $b_2$
\[ M_2^{(0)}\oplus K^{(0)}  = C^{-1}(b_2)=C^{-1}(b_1\oplus b) \]
Consequently, the initial key $K^{(0)}$ must satisfy
\begin{equation}
K^{(0)}  = C^{-1}(b_1)\oplus M_1^{(0)}= C^{-1}(b_2)\oplus M_2^{(0)}
\label{eq:c-diff5}
\end{equation}
If $C^{-1}(b_1)\oplus M_1^{(0)} \ne C^{-1}(b_2)\oplus M_2^{(0)}$, then
another choice of $b_1$ must be considered. Otherwise, the value of 
$K^{(0)}$ obtained from eq.~\ref{eq:c-diff5} is a possible key value.

The time complexity of this last step is thus $2^z$.

\subsection{Security and performance}\label{section:perf-security}

Let us now analyze in more detail eq.~\ref{eq:total-work}.
We write $N$ as  $N=2^\ell$, where $\ell=\log_2 N$. For $z=8=2^3$,
eq.~\ref{eq:total-work} becomes
\[ {\cal N}>2^{(\ell-3)r} 2^{(z-2)(r-d)} 2^{z-4}= 2^{(\ell+3)r-6d+4}
\]
where $r>d$ and $d=(1/2)\sqrt{N/z}=2^{(\ell-5)/2}$.
Some values of $N$, $r$ and ${\cal N}$ are shown in
table~\ref{table:1}.
Note that the size $N=32$ is only given to illustrate how $r$ changes when
the block size increases or decreases. Clearly, with 32-bit messages,
all messages $M$ could be generated to discover which one encrypts to
a given, intercepted $M'$. In this case, no complex cryptanalysis
would be necessary. 
\begin{table}
 \begin{center}
   \begin{tabular}{ccccc}
     $N$  & $\ell$  &  $d$  & $r$  & ${\cal N}$ \\
     \hline\\
     2048  & 11     &   8   & 13   & $2^{138}$ \\
      512  &  9     &   4   & 13   & $2^{136}$ \\
      128  &  7     &   2   & 14   & $2^{132}$ \\
       32  &  5     &   1   & 17   & $2^{134}$ \\
     \hline
   \end{tabular}
\end{center}
\caption{The work ${\cal N}$ needed to break the cipher as a function
  of block size $N$ and number of round $r$. The quantities $\ell$ and
  $d$ are by definition $\ell=\log_2 N$ and $d=(1/2)\sqrt{N/z}$. The
  number of bits per site is $z=8$.}
\label{table:1}
\end{table}

\begin{figure}
\begin{tabular}{m{.5\textwidth}m{.5\textwidth}}
\psfig{figure=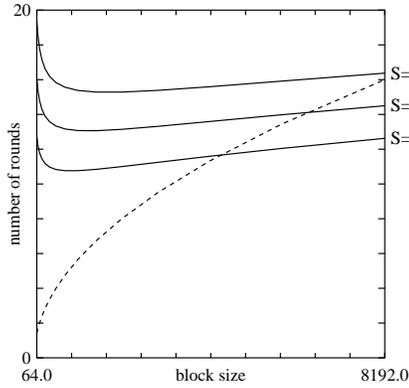,width=.45\textwidth} &
\psfig{figure=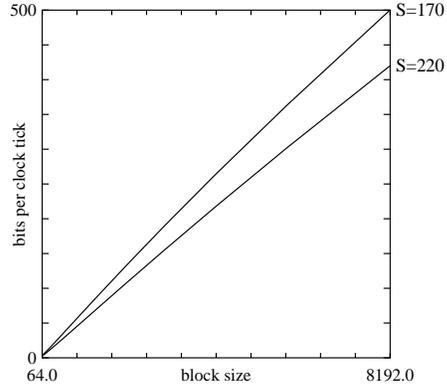,width=.45\textwidth}\\
\centerline{(a)} & \centerline{(b)} 
\end{tabular}
\caption{ (a) Number of rounds $r$ as a function of block size  $N$, to keep
a given security level $S$. Note that $r$ must be larger than the
diameter $d$. The limit $r=d$ is shown by the dashed curve. (b)
Throughput versus block size, for two given security levels and a
parallel implementation. The clock
tick is by definition the time required by one encryption round.}
\label{fig:round-versus-size}
\end{figure}

\begin{figure}
\begin{tabular}{m{.5\textwidth}m{.5\textwidth}}
\psfig{figure=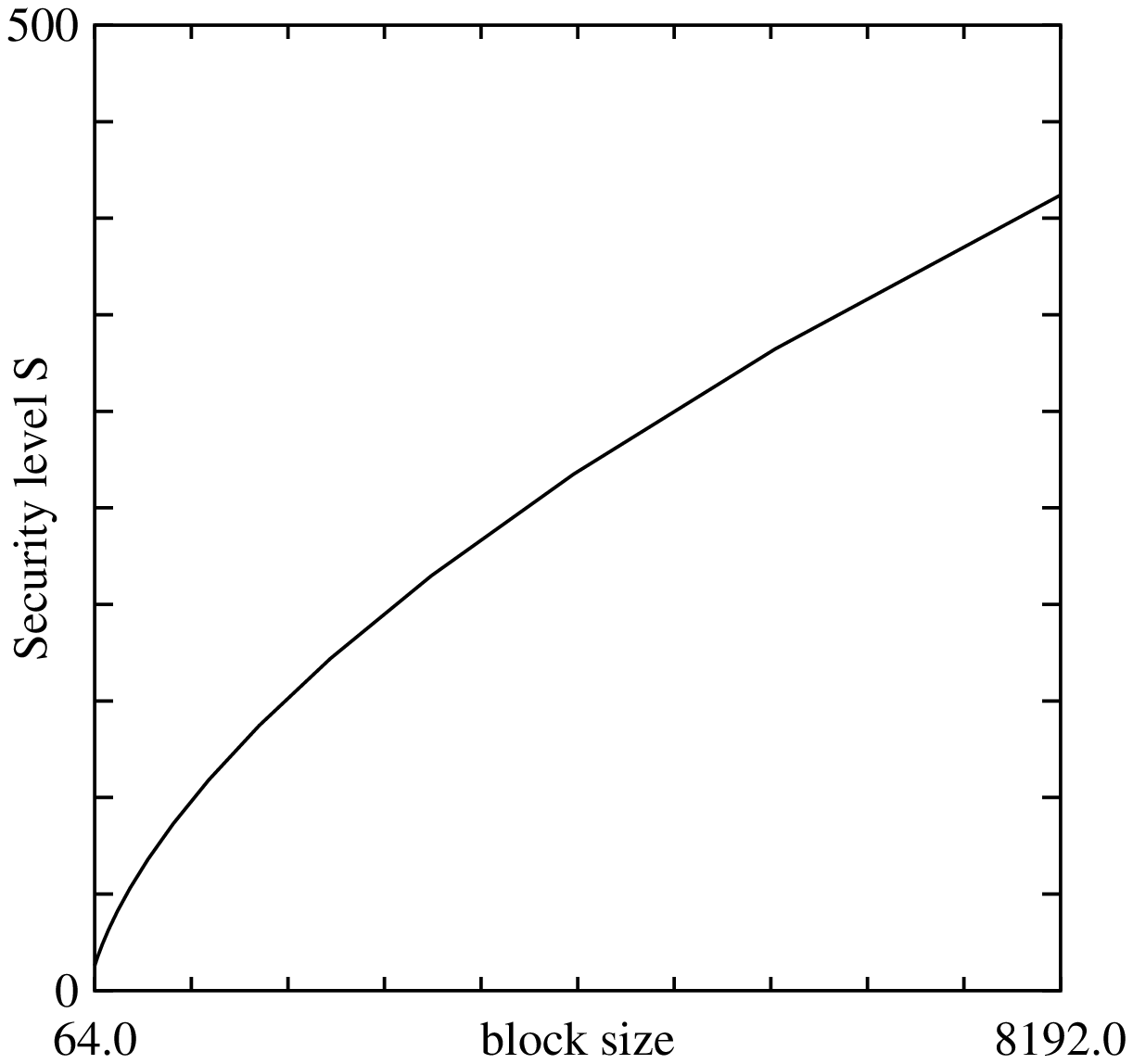,width=.45\textwidth} &
\psfig{figure=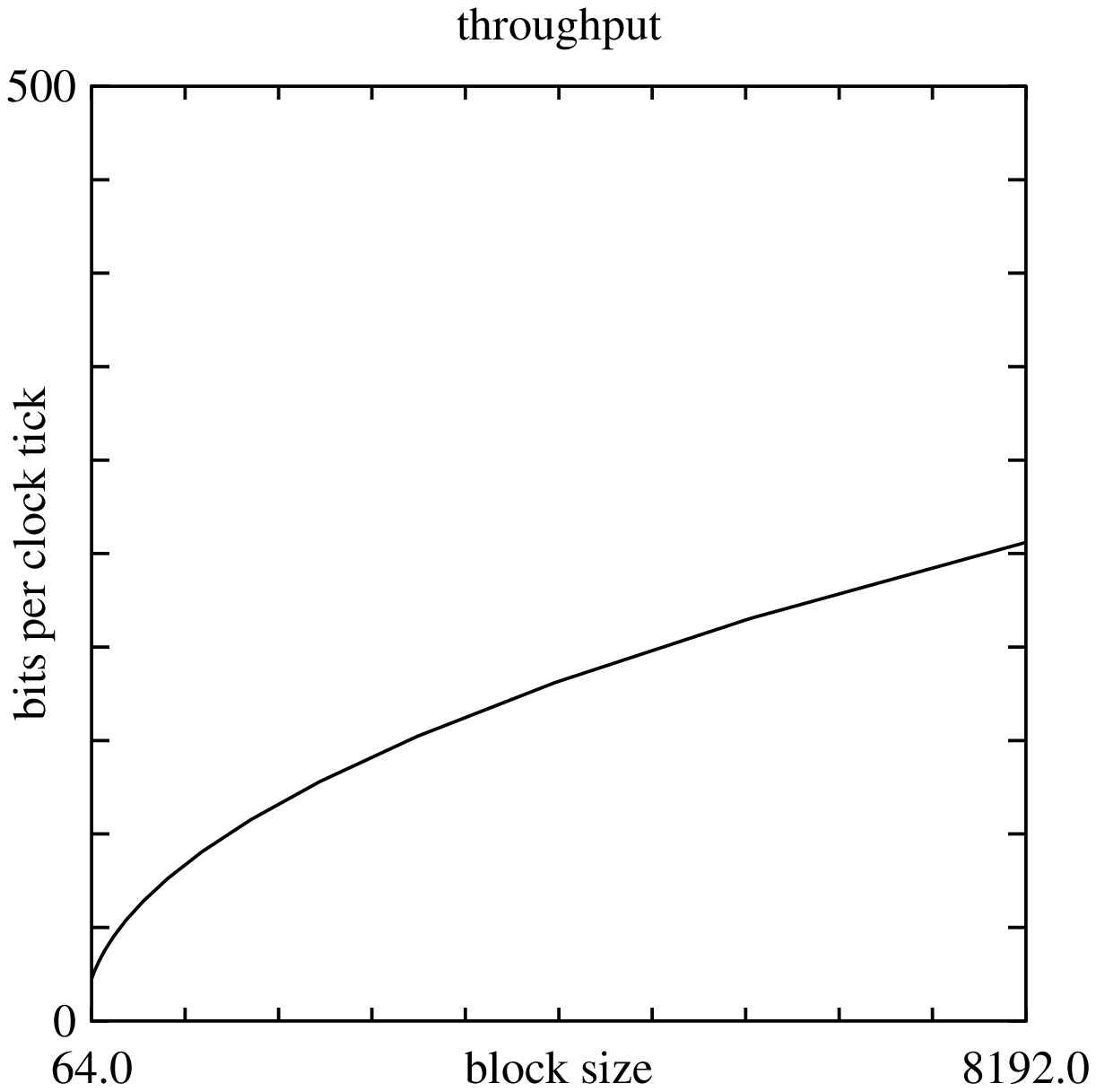,width=.45\textwidth}\\
\centerline{(a)} & \centerline{(b)} 
\end{tabular}
\caption{ (a) Security $S$ as a function of block size  $N$, 
for $r=2d$. (b) Throughput versus block size, for $r=2d$, in the case
of a parallel implementation. The clock
tick is time of one encryption round.}
\label{fig:security-versus-size}
\end{figure}

Some remarks about the throughput are now in order. With a parallel
hardware, the encryption time grows as the number of rounds but it is
independent of $N$. Thus, throughput clearly benefits from large $N$,
as the throughput is proportional to $N$ and inversely proportional to
$r$. In the examples shown in table~\ref{table:1}, we see that $N=512$
yields the best throughput while keeping about the same
cryptanalytic complexity.

This effect is mostly lost on a sequential hardware, since computing
time goes as $r\times N$ and the throughput is the block size divided
by the computing time. Thus, throughput is proportional to
$1/r$. Therefore, for a given value of ${\cal N}$, since $r$ increases
as $N$ decreases, larger sizes offer a better throughput.  However, it
may be argued that a software implementation of our algorithm on small
system size can be significantly faster. For instance, for $N=32$,
four table lookups of 256 entries are enough on a 32-bit architecture
to perform both the propagation and the collision on the entire
system. For $N=128$, 48 table lookups become necessary to implement at
once collision and propagation. Thus, for the same cryptanalytic complexity,
blocks of 32 bits would  provide a throughput about 10 times larger
than blocks of 128 bits.

In order to summarize the above discussion and to highlight the
link between security and performance in a {\bf parallel implementation},
let us define the security measure $ S$ as the logarithm of our estimate
of ${\cal N}$
\begin{equation}
{S}=\log_2 {\cal N} = (\ell+3)r -6d +4
\label{eq:security-measure}
\end{equation}
We also define the throughput per {\em clock tick}, $Q$,  as
\begin{equation}
Q={N\over r}={2^\ell \over r}
\end{equation}
where the {\em clock tick} is defined as the time needed to perform
one round of encryption.

In figure~\ref{fig:round-versus-size}~(a), we show how $r$ must change
with respect to $N$, for a given security level $S$.
We have from eq.~\ref{eq:security-measure}
\begin{equation}
 r={S+6d-4 \over \ell + 3} 
\label{eq:r-versus-S}
\end{equation}
so that, for each value of $N$ (i.e for the corresponding values of
$d$ and $\ell$), we can plot the value of $r$ ensuring the security
$S$. Note that, the constraint  $r>d$ must be satisfied.

From eq.~\ref{eq:r-versus-S} we also obtain the isosecurity
relation between $Q$ and $N$. This is shown in
figure~\ref{fig:round-versus-size}~(b). We see that for a clock rate
of 100 MHz, a throughput of 50Gb/s is achieved for a block of size
$N=8192$ bits.

Finally, in figure~\ref{fig:security-versus-size}~(a) we show how
security $S$ increases with $N$ when we take the number of round $r$
as twice the diameter $d$. In figure~\ref{fig:security-versus-size}~(b)
we plot $Q$, the resulting throughput per clock cycle.

\section{conclusion}

We have shown that discrete physics offers a natural and inspiring
framework to analyze symmetric cryptographic algorithms based of the
diffusion-confusion paradigm. Physical concepts such as entropy,
ergodicity, thermodynamic limits, mixing, Liapounov exponent, etc are
alternative ways to describe and quantify cryptographic devices and
security issues.  

The analogy with discrete physics provides us with simple design
principles to devise families of highly scalable encryption algorithms,
as discussed in detail in the present paper and exemplified with the
so-called \Crypto algorithm. A key feature of our approach is that
the encryption of large data blocks is possible with parallel hardware.
It results in a higher throughput and a higher security. In a time
where new applications develop and require the fast encryption of large
volume of data, our approach is a very promising solution to several
emerging technologies

\appendix
\section{Reversibility of the encryption-decryption algorithm}

We here prove that the decryption and encryption schemes are actually
identical.

The encryption algorithm consists of $n$ rounds of adding the key to
the message and performing a collision operation followed by a
propagation step. The last round terminates with the addition of the
current value of the key again. The key itself evolves simply with $n$
rounds of collision-propagation operations.

We propose to  write the full algorithm as
\begin{verbatim}
  reverse(M), reverse(K)
  propagation(M), propagation(K)
  repeat n times
    M=M xor K
    collision(M),  collision(K)
    propagation(M), propagation(K)
  end repeat
  M=M xor K
\end{verbatim}
As a consequence, the decryption will follow the exact same steps.
We sketch the proof below.

Let $M$ and $K$ be the initial message and key. After the first
application of reverse and propagation operator, it is convenient to define
\begin{equation}
M^{(0)}=PRM  \qquad K^{(0)}=K
\end{equation}

Let $M^{(m)}$ be the message after $m$ rounds and $K^{(m)}$ the
current key value at these $m$ rounds. With $P$ and $C$ the
propagation and collision operators, the above encryption rule reads
\begin{equation}
M^{(m)}= PC\left( M^{(m-1)} \oplus K^{(m-1)}\right) \qquad
   K^{(m)}= PC K^{(m-1)} 
\label{eq:reverse1}
\end{equation}
where $m=1,2,...,n-1$. After the last round, we add the key one more
time and we have
\begin{equation}
M^{(n)}= PC\left( M^{(n-1)} \oplus K^{(n-1)}\right) \oplus K^{(n)} \qquad
   K^{(n)}= PC K^{(n-1)} 
\label{eq:reverse2}
\end{equation}

In order to show that, by  applying the same steps a second time, the
message gets decrypted, we need to derive the following relations:
\begin{eqnarray}
PRM^{(n-1)} &=& PC\left(PRM^{(n)}\oplus PRK^{(n)}\right)  \oplus
PRK^{(n-1)} \label{eq:rev1} \\
PR M^{(m-1)} &=& PC PRM^{(m)} \oplus  PRK^{(m-1)} \qquad m=1,...,n-1 
\label{eq:rev2}\\
PR K^{(m-1)} &=&  PC PRK^{(m)} \qquad m=1,...,n  \label{eq:rev3}\\
\nonumber
\end{eqnarray}

The proof of the above three relations is now given.
From eq.~\ref{eq:reverse2} we obtain
\begin{equation}
RM^{(n)}\oplus RK^{(n)} = RPC\left( M^{(n-1)} \oplus K^{(n-1)}\right) 
\label{eq:reverse3}
\end{equation}
where $R$ is the reverse operator, which is linear: $R(a\oplus b)=Ra
\oplus Rb$.
By applying a propagation to both side of eq.~\ref{eq:reverse3} we obtain
\begin{equation}
PRM^{(n)}\oplus PRK^{(n)} = RC\left( M^{(n-1)} \oplus K^{(n-1)}\right) 
\label{eq:reverse4}
\end{equation}
because $P$ is linear and $PRP=R$. We can now apply $C$ on both side
and, since $CRC=R$, our equation becomes
\begin{equation}
C\left(PRM^{(n)}\oplus PRK^{(n)}\right) = RM^{(n-1)} \oplus RK^{(n-1)} 
\label{eq:reverse5}
\end{equation}
where, again we have used the linearity of $R$. This equation can be
rewritten as
\begin{equation}
RM^{(n-1)} = C\left(PRM^{(n)}\oplus PRK^{(n)}\right)  \oplus RK^{(n-1)} 
\label{eq:reverse6}
\end{equation}
or, after applying $P$ 
\begin{equation}
PRM^{(n-1)} = PC\left(PRM^{(n)}\oplus PRK^{(n)}\right)  \oplus PRK^{(n-1)} 
\label{eq:reverse7}
\end{equation}
This equation shows that $PRM^{(n-1)}$
is obtained from $PRM^{(n)}$ by the same expression as in
~\ref{eq:reverse2}. This proves  equation~\ref{eq:rev1}.

For the remaining $m<n-1$, we have, from eq.~\ref{eq:reverse1}
\begin{equation}
PRM^{(m)}= PRPC\left( M^{(m-1)} \oplus K^{(m-1)}\right) \qquad
   PRK^{(m)}= PRPC K^{(m-1)} 
\label{eq:reverse8b}
\end{equation}
or
\begin{equation}
PRM^{(m)}= RC\left( M^{(m-1)} \oplus K^{(m-1)}\right) \qquad
   PRK^{(m)}= RC K^{(m-1)} 
\label{eq:reverse9}
\end{equation}
which, by using $CRC=R$, can be further transformed into 
\begin{equation}
CPRM^{(m)}= R M^{(m-1)} \oplus RK^{(m-1)} \qquad
  CPRK^{(m)}= CRC K^{(m-1)} 
\label{eq:reverse10}
\end{equation}
or, also
\begin{equation}
PR M^{(m-1)}= PC PRM^{(m)} \oplus PRK^{(m-1)} \qquad
  PC PRK^{(m)}= PR K^{(m-1)} 
\label{eq:reverse11}
\end{equation}
These last equations prove eqs.~(\ref{eq:rev2}) and~(\ref{eq:rev3}).

Now we can apply our encryption algorithm a second time on $M^{(n)}$
and $K^{(n)}$ and show that it gives back $M$ and $K$. The result of
the first reverse and propagate operation gives $PRM^{(n)}$
and $PR K^{(n)}$. Now let us write the result of the first iteration.
Key addition plus collision and propagation yield
\[ PC\left(PR M^{(n)}\oplus PR K^{(n)} \right),  \qquad  PC PR K^{(n)} \]
Using relations~(\ref{eq:rev1}) and (\ref{eq:rev3}), this reduces to
\[ PR M^{(n-1)}\oplus PR K^{(n-1)},  \qquad  PR K^{(n-1)} \]

The next iteration starts by adding the current key, namely  $PR
K^{(n-1)}$. Thus the message now reads
\[ PR M^{(n-1)} \]
Then follows a collision-propagation step on both the message and the
key
\[ PC PR M^{(n-1)}, \qquad  PC PR K^{(n-1)} \]
Equations~(\ref{eq:rev2}) and (\ref{eq:rev3}) can now be applied for
$m=n-1$ and produce
\[ PR M^{(n-2)}\oplus PR K^{(n-2)},  \qquad  PR K^{(n-2)} \]
The next $n-2$ rounds work similarly and yield
\[  PR M^{(0)}\oplus PR K^{(0)},  \qquad  PR K^{(0)} \]
The final key addition (after the $n$ rounds) give
\[  PR M^{(0)}= PR PRM= R^2 M=M \]
This achieves the proof that the decryption is identical to the
encryption since the original message $M$ is recovered.

\section{Calculation of an inferior bound of $Q_r$}\label{appendix-2}

Let us define 
\[ Q_r=1\times3\times5...\times (2r+1) \]
This quantity can be written as 
\[ Q_r={(2r+1)! \over 2^r r!} \]
Thus, using Stirling formula, one has:
\begin{eqnarray}
\ln Q_r &=& \ln(2r+1)! - r\ln 2 - \ln r! \nonumber \\
         &=& (2r+1)\ln(2r+1) -(2r+1) - r\ln 2 - r\ln r + r \nonumber \\
       &\ge& 2r\ln(2r) -2r-1 - r\ln 2 - r\ln r + r \nonumber \\
         &=& 2r\ln2 + 2r\ln r -r-1 - r\ln 2 - r\ln r  \nonumber \\
         &=& r\ln2 + r\ln r - r - 1    \nonumber \\
         &=& r\ln2 + r\ln(2r/2) - r - 1    \nonumber \\
         &=& r\ln2 + r\ln2 + r\ln(r/2) - r - 1    \nonumber \\
         &=& r(2\ln2-1) + r\ln(r/2)  - 1    \nonumber \\
       &\ge& r\ln(r/2)  \nonumber \\
\end{eqnarray}
Therefore
\[ Q_r \ge \left( {r\over 2}\right)^r \]
and 
\[ Q_r^2 \ge \left( {r\over 2}\right)^{2r} \]
which is the quantity of interest to bound the amount of work for a
differential cryptanalysis of our algorithm.


\begin{thebibliography}{1}

\bibitem{BC-livre}
B.~Chopard and M.~Droz.
\newblock {\em Cellular Automata Modeling of Physical Systems}.
\newblock Cambridge University Press, 1998.

\bibitem{daemen:02}
Joan Daemen and Vincent Rijmen.
\newblock {\em The Design of Rijndael}.
\newblock Springer, 2002.

\bibitem{fhp-prl}
U.~Frisch, B.~Hasslacher, and Y.~Pomeau.
\newblock Lattice-gas automata for the navier-stokes equation.
\newblock {\em Phys. Rev. Lett.}, 56:1505, 1986.

\bibitem{marsaglia:85}
A.~Current G.~Marsaglia.
\newblock View of random number generators.
\newblock In L.~Billard, editor, {\em Computer Science and Statistics, The
  Interface}. Elsevier Science, 1985.

\bibitem{kirkpatrick:81}
S.~Kirkpatrick and E.P. Stoll.
\newblock A very fast shift-register sequence random number generator.
\newblock {\em Journal of Computational Physics}, 40:517--526, 1981.

\bibitem{pourLaScience}
Pour la~Science: dossier~hors série, editor.
\newblock {\em L'Art du Secret}, 2002.

\bibitem{schneier:96}
Bruce Schneier.
\newblock {\em Applied Cryptography}.
\newblock Wiley, 1996.

\bibitem{shannon:48}
C.~E. Shannon.
\newblock Mathematical theory of communication.
\newblock {\em Bell Syst. Tech. Journal}, 27(3):379--423 and 623--656, 1948.

\bibitem{shannon:49}
C.~E. Shannon.
\newblock Commuication theory of secrecy systems.
\newblock {\em Bell Syst. Tech. Journal}, 28:656--715, 1949.

\end{thebibliography}

\end{document}